\documentclass[11pt]{article}
\usepackage{yandex-research}

\newcommand{\Description}[1]{}

\usepackage{amsmath}
\usepackage{booktabs}
\usepackage{graphicx}
\usepackage{tabularx}
\usepackage[english]{babel}

\title{X-split Argus: Learning Reusable History Tokens for Efficient CTR Prediction}

\author{%
  \fontsize{10}{14}\selectfont
  Alexander Ploshkin\textsuperscript{*},
  Nikita Stepanov\textsuperscript{*},
  Alexey Pismenny\textsuperscript{*}\\
  Viacheslav Ovchinnikov, Nikolay Savushkin
}
\affiliations{}
\date{}
\hypersetup{
  pdftitle={X-split Argus: Learning Reusable History Tokens for Efficient CTR Prediction},
  pdfauthor={Alexander Ploshkin, Nikita Stepanov, Alexey Pismenny, Viacheslav Ovchinnikov, Nikolay Savushkin}
}

\begin{document}
\maketitle
\begingroup
\renewcommand{\thefootnote}{\fnsymbol{footnote}}
\footnotetext[1]{These authors contributed equally to this work}
\endgroup
\begin{abstract}
Accurate click-through rate (CTR) prediction depends on user representations
that capture both persistent preferences and recent intent, yet processing
extensive behavioral histories for every ranking request is costly under
production compute and latency constraints. We present X-split ARGUS, a
framework for ad ranking whose history compressor and online ranker are
trained end to end. The compressor encodes past interactions into a compact
set of tokens that are computed offline and reused across candidates, while
the online ranker scores ads through cross-at\-ten\-tion to these tokens and
representations of recent events. Training combines an autoregressive loss for
click feedback with distillation of candidate-specific click logits from a
larger teacher that scores each candidate using the uncompressed history. The
number of compressed history tokens and the length of recent history can be
adjusted to fit serving compute and latency budgets. Offline evaluation shows
a 0.81\% relative reduction in normalized entropy (NE) compared with the
production baseline. Deployment in a large-scale real-time bidding system
increases online inference cost by 15\% relative to the production baseline.
An online A/B test against this baseline yields relative gains of
\textbf{4.65\%} in a business metric reflecting the monetary value of
conversions generated for advertisers and \textbf{2.96\%} in clicks.
\end{abstract}




\section{Introduction}
\label{sec:introduction}

Accurate click-through rate (CTR) prediction requires user representations that
capture persistent preferences, respond to recent intent, and retain behavioral
signals relevant to each candidate ad. Longer histories can provide additional
evidence for these predictions, and recent sequential models demonstrate the
benefits of scaling behavioral context and model capacity~\cite{hstu,kunlun}.
However, encoding extensive histories for each ranking request increases GPU
computation and memory traffic, while processing that depends on the candidate can limit reuse across ads. At production request rates, these costs must fit strict
compute and response latency budgets, making the representation of historical
context a central deployment decision.

Existing systems already address important parts of this problem.
TransAct~\cite{transact} and LLaTTE~\cite{llatte} combine precomputed user
representations with online processing of recent behavior, while
VISTA~\cite{vista} learns reusable history summaries for candidate-aware
scoring. TM20K~\cite{tm20k} further combines sequence compression with
supervision from a teacher that retains the full history. These approaches
motivate a concrete question for our deployment: how much of a teacher's predictive benefit from uncompressed history can be retained when past interactions are represented by a compact token set that does not depend on the candidate?
We study this question for a jointly trained compressor and ranker, with
explicit control over the history processed online.

We present \textbf{X-split ARGUS}, a framework that uses autoregressive
feedback prediction, as in ARGUS~\cite{argus}, while changing the
interface between history modeling and candidate scoring. ARGUS uses
two-tower fine-tuning to support scoring with precomputed user and item
representations. In X-split ARGUS, a history compressor instead produces a set
of tokens without conditioning on the candidate, and an online ranker scores
ads through cross-at\-ten\-tion to these tokens and representations of recent
events. The compressor and ranker are trained end to end, while the history
tokens are precomputed offline for reuse during serving. Their number and the
length of recent history determine the amount of history processed online.

Observed click feedback and click logits for individual candidates from a
larger teacher jointly supervise the compressor and online ranker. The teacher
scores ads using the uncompressed history. The distillation objective reaches the history
compressor through the online ranker, allowing observed feedback and teacher
predictions to shape the information retained in the summaries.

For online evaluation, we use \emph{Ad Conversion Value} (ACV), our primary
internal metric for accepting advertising A/B experiments. ACV reflects
the monetary value of advertiser conversions delivered at the same advertising
budget, capturing the benefit of improved conversion delivery.

Our main contributions are as follows:
\begin{itemize}
    \item \textbf{Joint offline--online history modeling.}
    We introduce a jointly trained history compressor and candidate-aware
    ranker that combines reusable history tokens with recent events.
    The number of summary tokens and the length of recent history control the amount of history processed online.

    \item \textbf{Training and evaluation under history compression.}
    We combine autoregressive prediction losses with supervision from the click logits of a teacher using uncompressed history and evaluate the predictive quality retained by the compressed student. In the main offline comparison,
    X-split ARGUS reduces normalized entropy (NE) by 0.81\% relative to
    Context-aware ARGUS and retains over 70\% of the teacher's improvement
    over the same baseline.

    \item \textbf{Production deployment and measured impact.}
    We deploy X-split ARGUS on an advertising surface of our real-time bidding
    system handling roughly a hundred thousand requests per second.
    The model meets the existing latency budget using 15\% more online
    GPUs of the same type to serve the same traffic. An online A/B test
    against the production baseline yields relative gains of \textbf{4.65\%} in
    ACV and \textbf{2.96\%} in clicks.
\end{itemize}

\section{Related Work}
\label{sec:related_work}

\subsection{Sequential Modeling in Recommender Systems}
\label{sec:rw_sequential}

Early attention-based sequential recommenders established complementary
learning objectives: SASRec~\cite{sasrec} predicts the next item using causal
self-at\-ten\-tion, BERT4Rec~\cite{bert4rec} reconstructs masked items from
bidirectional context, and S$^3$-Rec~\cite{s3rec} learns associations among
attributes, items, and subsequences through self-su\-per\-vised pretraining.
For CTR prediction, DIN~\cite{din} makes history aggregation dependent on the
candidate through a local activation mechanism. This distinction between
learning a reusable user state and extracting information for a particular
candidate remains central to modern ranking architectures.

HSTU~\cite{hstu} scales retrieval and ranking as sequential transduction using
pointwise aggregated attention and multiplicative feature transformations.
ARGUS~\cite{argus} combines autoregressive feedback and next-item prediction
with impression-aware two-tower fine-tuning for offline representation
computation. X-split ARGUS exposes multiple history tokens to an online ranker
that conditions its predictions on each candidate. Kunlun~\cite{kunlun}
addresses compute allocation through hierarchical seed pooling and
\emph{Computation Skip}, which alternates self-at\-ten\-tion and summarization
modules across layers.

Industrial rankers also differ in how they combine history and nonsequential
features. WHALE~\cite{whale} connects Wukong feature interactions to HSTU
sequence states through attention; MTGR~\cite{mtgr} retains engineered cross
features in candidate tokens within an encoder based on HSTU; and
OneTrans~\cite{onetrans} processes sequential and nonsequential tokens in a
unified backbone. HyFormer~\cite{hyformer} alternates attention from global
queries to history representations with interactions among the queries.
These designs concern the interaction between history and ranking features;
the timing and reuse of history computation introduce an additional serving
choice.

Sequential modeling also supports generative retrieval, which predicts
item identifiers from user histories. TIGER~\cite{tiger}
generates semantic item identifiers, while Gryphon~\cite{gryphon} shares a
history encoder between identifier generation and scoring individual items.
These methods illustrate the use of autoregressive history representations
across retrieval and ranking; our focus is the prediction of click feedback
for supplied candidate ads under online serving constraints.

\subsection{Efficient Modeling of Long Histories}
\label{sec:rw_compression}

Systems for long histories differ in what they retain, when they compute it, and how
candidates access it. Event selection, representation compression, and
computation reuse can coexist within the same architecture. Particularly
relevant to our setting, TransAct~\cite{transact} combines user embeddings computed in batches with a Transformer conditioned on the candidate to process recent actions.
LLaTTE~\cite{llatte} scales an asynchronous user model whose cached
representations augment a compact online ranker with fresh events and ad
context. Both establish the use of separate paths for historical context and
recent intent.

VISTA~\cite{vista} learns summary tokens for downstream candidate attention
and caches these tokens for reuse. Its reported variants include joint
training with HSTU and an auxiliary objective for sequence reconstruction.
SITA~\cite{sita} organizes stored interest tokens through semantic identifiers
and uses each candidate's identifiers to select tokens before target attention.
Our design uses cross-at\-ten\-tion from candidates to learned history summaries and representations of recent events. A teacher using uncompressed history supervises the jointly trained compressor and ranker. We study the quality
retained by this combination under production constraints.

Systems using retrieval retain selected events or historical embeddings.
SIM~\cite{sim} retrieves a subsequence relevant to the candidate before applying
finer target attention, and TWIN~\cite{twin} aligns the relevance measures of
these two stages. TransAct~V2~\cite{transactv2} selects nearest neighbors from
lifelong, recent, and impression histories while explicitly keeping the
freshest actions. Representation Memento~\cite{memento} retrieves relevant
and diverse temporally aggregated embeddings using a modified maximal
marginal relevance criterion. These approaches select context according to the candidate, controlling what is passed to the ranker.

Other methods reduce the history representation through aggregation.
TWIN~V2~\cite{twinv2} uses offline hierarchical clustering and cluster-aware
target attention; CAUSE~\cite{cause} forms history tokens by category while
retaining recent events; and LONGER~\cite{longer} combines local token merging
with cross-at\-ten\-tion into a shorter sequence. More general attention
bottlenecks, including Perceiver's latent array~\cite{perceiver} and the learned
queries of BLIP-2's Q-Former~\cite{blip2}, provide architectural precedents for
extracting compact representations from larger inputs.

Representation reduction and computation reuse also interact within serving
requests. HELIX~\cite{helix} constructs compressed history states independently of the candidate once per request and lets tokens conditioned on each candidate repeatedly attend to them. This execution strategy differs from computing
summaries of long histories offline, as in our deployment.
\emph{Make It Long, Keep It Fast}~\cite{makelong} uses stacked cross-at\-ten\-tion from targets to history and batching within each request to share history encoding;
OneTrans~\cite{onetrans} reuses causal sequence computations; and
MARM~\cite{marm} caches intermediate history states for multilayer target
attention. These reuse strategies reduce repeated computation without
necessarily imposing a compact summary representation.

\subsection{Distillation under Serving Constraints}
\label{sec:rw_distillation}

Knowledge distillation supplements observed labels with teacher predictions
to train a constrained student~\cite{hintonkd}. Ranking
Distillation~\cite{rankingdistillation} transfers a teacher's top-ranked
unobserved items as weighted supervision, while Collaborative
Distillation~\cite{collaborativedistillation} combines soft targets with
rank-aware sampling. Privileged Features
Distillation~\cite{privilegeddistillation} addresses a different asymmetry:
teachers can use predictive features unavailable to the deployed student.
In our setting, historical interactions remain available to the offline
compressor; the constraint concerns their representation and processing during
online ranking.

TM20K~\cite{tm20k} is a close precedent for distillation under history
compression: a teacher retains the full sequence, while students merge tokens
according to item identity, position, and layer depth. X-split ARGUS learns
summary tokens for offline precomputation and reuse alongside recent events.
Gryphon-v2~\cite{gryphonv2} distills scores for generated and logged candidates
into both a ranking module and its shared history encoder.
Meta Lattice~\cite{metalattice} combines logit supervision with asynchronously
precomputed teacher embeddings for user--item pairs, which students consume
during both training and serving. Our stored summaries do not depend on the candidate; teacher logits for individual candidates supervise their learning through the online ranker. The focus of our evaluation is the predictive quality retained by this
representation and training design under production serving constraints.

\newpage
\section{X-split ARGUS Design}
\label{sec:design}

X-split ARGUS combines precomputed historical context with online modeling of
recent behavior, while allowing supervision for individual candidates to train both
parts end to end. Let $H=(e_1,\ldots,e_n)$ be a chronological user history with
$n\leq N$, where $e_t=(x_t,y_t)$ contains item features $x_t$ and observed
binary click feedback $y_t$. The item features comprise an item identifier and
tokenized metadata. For a target event at position $t$, the model predicts
$y_t$ using $x_t$ and information available before that event. We use $M$
representations of recent events and $K$ compressed history tokens as the
candidate's historical memory. The student consists of a history compressor,
a runtime user encoder, a candidate encoder, and a decoder with cross-at\-ten\-tion,
as shown in Figure~\ref{fig:x_split_architecture}. A teacher that processes
the history without compression provides additional training supervision.
Numerical hyperparameters are reported in
Appendix~\ref{app:design_configuration}.
Throughout this section, $[\cdot;\cdot]$ denotes concatenation along the
sequence dimension, while $[\cdot\Vert\cdot]$ denotes concatenation along the
feature dimension. We omit model parameter subscripts for readability.

\begin{center}
\begin{minipage}{\linewidth}
    \captionsetup{type=figure}
    \centering
    \includegraphics[width=0.85\linewidth]{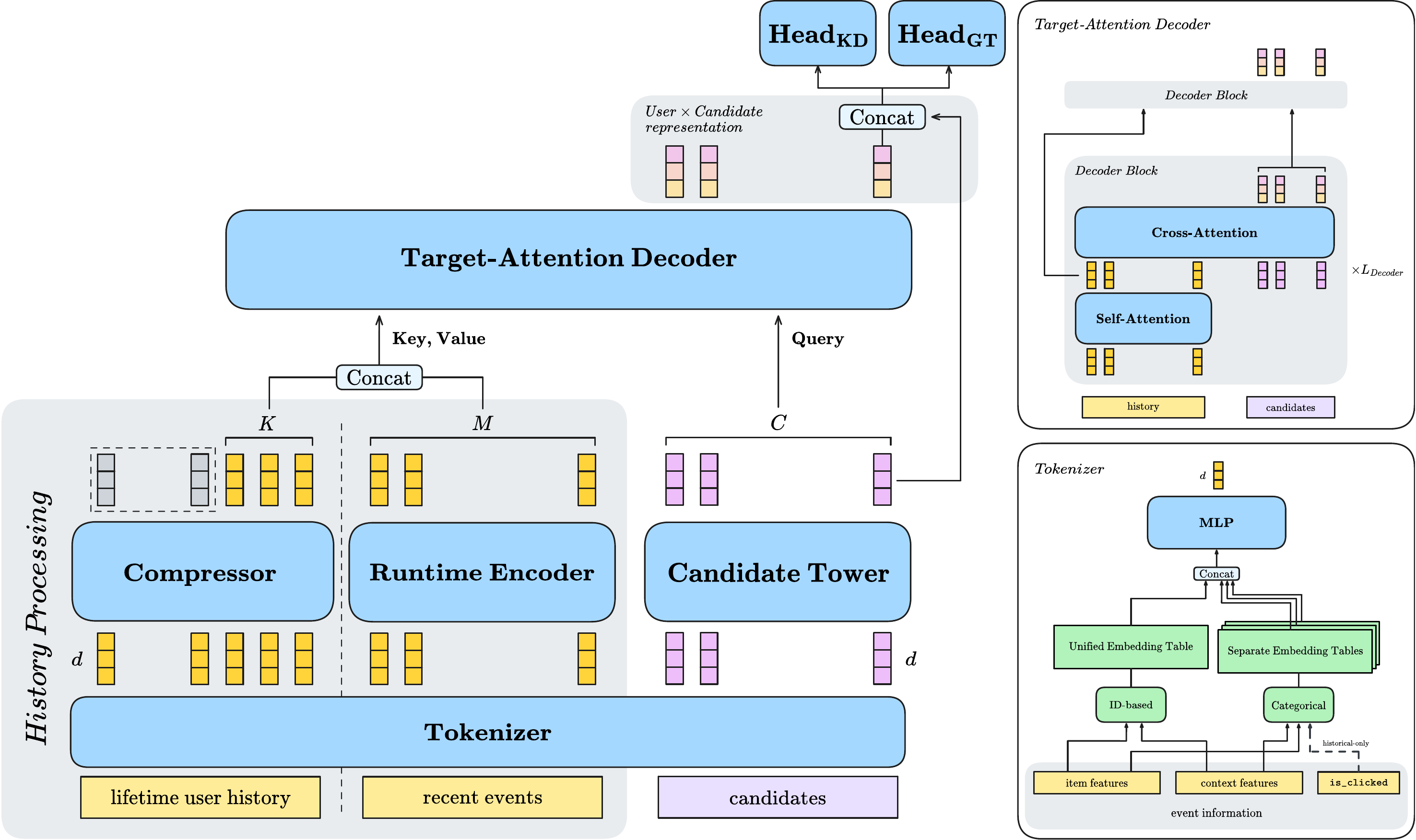}
    \caption{X-split ARGUS architecture: jointly trained history compression and
    candidate-aware ranking with recent events, supervised click prediction,
    and knowledge distillation.}
    \Description{The architecture combines a tokenizer, a history compressor, a runtime encoder, a candidate encoder, and a target-attention decoder. The tokenizer uses a unified embedding table for ID features and separate tables for categorical features, concatenates the embeddings, and applies an MLP. Click feedback is included only for historical events. The compressor and runtime encoder supply a shared memory of historical states. Each decoder block applies self-attention only to this history and cross-attention from candidate queries. Decoder outputs are concatenated with candidate representations for separate supervised and distillation heads; the supervised head supplies serving predictions.}
    \label{fig:x_split_architecture}
\end{minipage}
\end{center}

\subsection{Student Architecture and Serving}
\label{sec:design_student}

\paragraph{History compressor.}
The compressor is a Transformer~\cite{transformer} with causal self-at\-ten\-tion
over historical events, including their observed feedback. It produces one
contextualized state per input event:
\begin{equation}
    \begin{gathered}
        (c_1,\ldots,c_n)=\mathcal C(H),\\
        c_i\text{ depends only on }e_1,\ldots,e_i.
    \end{gathered}
    \label{eq:compressor_states}
\end{equation}
For a prefix ending at event $b$, where $K\leq b\leq n$, its compressed
representation consists of the last $K$ states of that prefix:
\begin{equation}
    Z(b)=[c_{b-K+1};\ldots;c_b].
    \label{eq:compressed_prefix}
\end{equation}
Selecting the final $K$ contextualized states gives a compact history
representation while allowing compressed representations for multiple
training cutoffs to be extracted from a single causal pass. Each retained
state carries information from its causal prefix, and the final state can
attend to every event up to $b$. The student prediction objectives train
these states to preserve information useful for subsequent candidate scoring.
Their computation does not depend on the candidates scored against them.

\paragraph{Runtime user encoder.}
The runtime encoder processes recent events and their observed feedback using
causal self-at\-ten\-tion with a sliding window. The window at each layer controls
the temporal receptive field of the output states; stacked layers with local attention
can propagate information beyond a single layer's window~\cite{longformer}.
For a serving request, let $H^{\mathrm{recent}}$ contain up to the most
recent $M$ events available in the user profile. The online representations are
\begin{equation}
    U=E(H^{\mathrm{recent}}).
    \label{eq:runtime_serving}
\end{equation}
The encoder uses neither candidate features nor decoder outputs. The use of a
shared sequence pass during training is described in
Section~\ref{sec:design_training}.

\paragraph{Candidate encoder and decoder.}
\label{sec:design_decoder}
The candidate encoder maps an ad's item identifier and metadata to a
single item token $q_a=A(x_a)$, which excludes the candidate's
click outcome. The student decoder receives this token without
introducing a separate feedback token. It conditions the item
representation on both sources of user context through cross-at\-ten\-tion, with
candidate states supplying queries and history states supplying keys and
values~\cite{transformer}. Its historical memory is
\begin{equation}
    V=[Z(b);U],
    \label{eq:serving_memory}
\end{equation}
where $b$ is the cutoff of the cached history representation available for
the request. The two history streams have the same feature dimension for
this concatenation. Thus, the compressed tokens join the states of recent
events \emph{after} the runtime
encoder, and the decoder attends to $K+M$ history states for a full window.

The optional decoder self-at\-ten\-tion (SA) operates only over these $K+M$
history states, using full attention within the historical memory.
Candidate tokens do not participate in this operation, so the resulting
history representations remain independent of the candidates being scored.
All states in the memory precede the prediction target; allowing them to
attend to one another does not expose the target's feedback or future events.
Candidate scores remain independent when evaluated against the same memory.

As shown in Figure~\ref{fig:x_split_architecture}, a second concatenation
combines the decoder output $d_a=D(q_a,V)$ with the candidate
representation along the feature dimension, yielding $r_a=[d_a\Vert q_a]$.
This gives both prediction heads access to the representation conditioned
on user history and the original candidate representation. The supervised
head produces the serving CTR prediction.

\paragraph{Computation reuse.}
The architecture supports three scopes of execution
(Figure~\ref{fig:x_split_architecture}). The compressor outputs are precomputed
and cached across requests. Within each request, the runtime encoder and
any self-at\-ten\-tion over the decoder's historical memory are independent of
the candidates and can be shared across them. Candidate encoding,
cross-at\-ten\-tion to this memory, and the prediction heads provide computation
for individual candidates. The absence of a dependency from candidate
scoring back to either history encoder enables this reuse. The budgets $K$
and $M$ control the decoder memory size, while $M$ also controls how many raw
events are encoded online.

\paragraph{Snapshot and recent history boundaries.}
For a request with $n$ available historical events, let $s=\max(0,n-M)$.
The runtime input contains events $s+1,\ldots,n$, whereas the cached
compressor states summarize a prefix ending at $b\leq n$.
These boundaries need not coincide: $b=s$ gives adjacent coverage,
$b<s$ leaves events $b+1,\ldots,s$ outside both branches, and
$b>s$ represents events $s+1,\ldots,b$ in both branches.
This distinction concerns which events are available to the two branches;
it is separate from the difference in the runtime encoder's receptive field
between training and serving, described in Section~\ref{sec:design_training}.

\subsection{Teacher with Uncompressed History}
\label{sec:design_teacher}

The teacher is a larger Transformer that processes the retained event history
$H$ with full causal self-at\-ten\-tion, without compressing it into summary
states. For a target at position $t$, it uses the current item features $x_t$
and the same retained events preceding $t$; its access to these events is
not restricted to the student's $K$ compressed states and $M$ recent states.
It follows the
autoregressive formulation of feedback prediction used in HSTU and
ARGUS~\cite{hstu,argus}: item information precedes its observed feedback,
allowing the model to predict the latter from the former and the preceding
history. In our teacher, each target event is represented by an item token
followed by a feedback token. The item token contains the item features but
excludes $y_t$; the feedback token carries the observed click
outcome. This separation into two tokens is specific to the teacher. The
student decoder uses the single item token described above, while both
student history encoders receive observed feedback as an event feature.

Let $\widetilde H$ denote this token sequence and $\pi(t)$ the position of
the item token for target event $t$. The teacher's click logit is
\begin{equation}
    \begin{aligned}
        (v_1^T,\ldots,v_{|\widetilde H|}^T)&=T(\widetilde H),\\
        z_t^T&=g_T(v_{\pi(t)}^T).
    \end{aligned}
    \label{eq:teacher_logit}
\end{equation}
Here $v_j^T$ is the teacher state at token position $j$, $g_T$ is its click
prediction head, and the superscript $T$ identifies teacher quantities.
The feedback token for event $t$ occurs after $\pi(t)$ and is unavailable to
its own click prediction. Once present in the prefix, the event and its
feedback can contribute to subsequent predictions. Teacher and student
outputs are aligned by target event rather than by token index, since their
sequence representations differ. The teacher is trained separately from the
student, and its checkpoint provides fixed logit targets during student
updates. No student gradient updates the teacher, which is absent from the
online scoring path.

\subsection{End-to-End Autoregressive Training}
\label{sec:design_training}

\paragraph{Sequence construction and temporal boundaries.}
We train on complete retained user histories. Let $\mathcal T(H)$ denote
the set of event positions selected for click supervision in history $H$.
For a target event $t\in\mathcal T(H)$, the student decoder
receives an item token computed from $x_t$, and $y_t$ serves only as the
prediction target. Historical events supplied to the compressor and runtime
encoder include their observed feedback, but the decoder memory for event
$t$ is assembled exclusively from states preceding $t$.

For the default construction at target position $t$, the boundary separating
the selected recent events from the older prefix is
\begin{equation}
    b(t)=\max(0,t-1-M).
    \label{eq:training_cutoff}
\end{equation}
This training boundary is defined for each target; the cached cutoff $b$
used at serving time is determined by the available offline snapshot and
need not follow this rule. Both history encoders process the sequence in
a single causal pass. Writing $(u_1,\ldots,u_n)=E(H)$, for a full memory
window with $b(t)\geq K$ the target's decoder memory is assembled as
\begin{equation}
    \begin{aligned}
        U_t&=[u_{b(t)+1};\ldots;u_{t-1}],\\
        V_t&=[Z(b(t));U_t].
    \end{aligned}
    \label{eq:training_memory}
\end{equation}
All selected states precede the target event. Consequently, neither its
feedback nor future events enter its click prediction through the history
branches. The input event sequence is shared across prediction positions,
avoiding a separate pass over the historical prefix for every target.
Each supervised event contributes a prediction target at its own position
and historical context at later positions.
In the history budget experiments of Section~\ref{sec:offline_ablations},
we hold the boundary at its reference value while varying the numbers
of retained compressor and runtime states.

\paragraph{Runtime context during training and serving.}
Independently of the cached prefix boundary, the runtime encoder's shared
training pass and its truncated serving input
have different left boundaries. During training, early states within the
selected recent block can incorporate older events through local attention.
During serving, only the most recent $M$ raw events are supplied to the
encoder, truncating this preceding context. Thus, $M$ specifies the selected
runtime states during training and the raw online input length during
serving; it does not impose an identical total receptive field in both modes.

\paragraph{Click prediction.}
For target event $t$, let $q_t=A(x_t)$ and
$d_t=D(q_t,V_t)$. The two prediction heads receive
the shared representation $r_t=[d_t\Vert q_t]$ shown in
Figure~\ref{fig:x_split_architecture}. The supervised head
produces $z_t^{\mathrm{GT}}=g_{\mathrm{GT}}(r_t)$ and
$p_t=\sigma(z_t^{\mathrm{GT}})$, where $\sigma$ is the logistic sigmoid.
The binary cross-entropy for this event is
\begin{equation}
    \begin{aligned}
        \ell_{\mathrm{GT},t}={}&-y_t\log p_t\\
        &-(1-y_t)\log(1-p_t).
    \end{aligned}
    \label{eq:click_loss}
\end{equation}

\paragraph{Optional smoothing of negative targets.}
\label{sec:design_smoothing}
Some configurations regularize the supervised objective with asymmetric
label smoothing~\cite{inception2016,labelsmoothing2019}. The training target
for an impression without a click is raised from zero to $\varepsilon\in[0,1)$,
while an observed click retains a target of one:
\begin{equation}
    \widetilde y_t=y_t+\varepsilon(1-y_t).
    \label{eq:negative_target_smoothing}
\end{equation}
For these configurations, $\widetilde y_t$ replaces $y_t$ in
Equation~\ref{eq:click_loss}. Smoothing applies only to the supervised loss
and leaves the teacher logits used for distillation unchanged. The value
of $\varepsilon$ is reported in Appendix~\ref{app:design_configuration}.

This regularization is motivated by the limited evidence about user
interest provided by an impression without a click: the user may not have
seen the ad. An unclicked impression remains a valid negative outcome for
CTR prediction; smoothing softens the training objective without treating
that observation as an incorrect label. We evaluate this optional training
choice in Section~\ref{sec:offline_smoothing}.

\paragraph{Knowledge distillation.}
Teacher supervision supplements observed clicks through knowledge
distillation~\cite{hintonkd}. We use separate supervised and distillation
heads over the same student representation, an arrangement also explored in
distillation for industrial sequence models~\cite{tm20k}. The auxiliary head
produces $z_t^{\mathrm{KD}}=g_{\mathrm{KD}}(r_t)$ and matches the teacher's
raw click logit using an $L_1$ objective:
\begin{equation}
    \ell_{\mathrm{KD},t}
    =\left|z_t^{\mathrm{KD}}-
      \operatorname{stopgrad}(z_t^T)\right|.
    \label{eq:distillation_loss}
\end{equation}
The $L_1$ loss penalizes discrepancies between raw student and teacher
logits linearly. The supervised head is trained with the click prediction
loss, while the auxiliary distillation head is trained to match teacher
logits. Both losses update the shared student components, allowing teacher
supervision to shape the representations used by the serving head.
At serving time, predictions are obtained exclusively from
$g_{\mathrm{GT}}$; the distillation head is used only during training.

\paragraph{Joint objective and parameter updates.}
Let $\mathcal L_{\mathrm{GT}}$ and $\mathcal L_{\mathrm{KD}}$ denote the
corresponding aggregate losses over positions in $\mathcal T(H)$. The combined
click supervision and distillation objective is
\begin{equation}
    \mathcal L
    =\mathcal L_{\mathrm{GT}}
    +\lambda_{\mathrm{KD}}\mathcal L_{\mathrm{KD}}.
    \label{eq:joint_objective}
\end{equation}
The nonnegative weight $\lambda_{\mathrm{KD}}$ controls the contribution of
distillation and is specified in Appendix~\ref{app:design_configuration}.
For each training sequence, we obtain teacher logits for the target events,
compute both student history streams, select the prefix and recent states
for each target, and evaluate the decoder objectives. The student components
and prediction heads are optimized jointly. In particular, gradients of the
student prediction objectives propagate through the decoder into both the
runtime encoder and the history compressor. Teacher outputs remain detached
during these updates. Offline precomputation is therefore an execution
choice for serving, while the compressor remains part of the differentiable
student during training.

\section{Offline Evaluation}
\label{sec:offline_evaluation}

\subsection{Experimental Setup and Metrics}
\label{sec:offline_metric}

\paragraph{Dataset and global temporal split.}
We use a large internal industrial dataset containing all data collected by
our real-time bidding system. Table~\ref{tab:offline_dataset} summarizes the
dataset, retained user histories, and temporal split. We use the same
training and test time boundaries across users. The gap between the two
periods emulates the delay between the completion of model training and
the model's actual deployment.

\begin{table}[tbp]
    \centering
    \caption{Dataset and temporal split. History lengths count original
    events; the cap corresponds to the 99.5th percentile of user event counts.}
    \label{tab:offline_dataset}
    \small
    \setlength{\tabcolsep}{4pt}
    \begin{tabularx}{\linewidth}{@{}>{\raggedright\arraybackslash}X>{\raggedright\arraybackslash}X@{}}
        \toprule
        Parameter & Value \\
        \midrule
        Source & Internal RTB logs \\
        Data and history horizon & 365 days \\
        Training unit & User history \\
        Training targets (including repeats) & $\sim 10^9$ \\
        History cap & $8{,}192$ events \\
        Target & Binary click feedback \\
        Split & Global temporal \\
        Train--test gap & 1 day \\
        Test period & 3 days \\
        \bottomrule
    \end{tabularx}
\end{table}

\paragraph{Prediction context and model checkpoints.}
Each model's checkpoint remains fixed throughout the test period,
while both compressed history and runtime history are updated as evaluation
progresses. Context uses events available before the impression being
scored, including events from the gap and the elapsed portion of the test
period. For X-split ARGUS predictions, the runtime encoder is recomputed on
the last $M$ raw events preceding the target. Its states are obtained from
the restricted recent input, rather than selected from a shared pass over
the full history. All configurations are evaluated on the same set of
target impressions.

\paragraph{Evaluated configurations.}
We compare X-split ARGUS with Context-aware ARGUS, described in
Section~\ref{sec:offline_baseline}, and the uncompressed teacher from
Section~\ref{sec:design_teacher}. The X-split ARGUS configuration in this
comparison uses target attention, knowledge distillation, self-at\-ten\-tion
over the decoder's historical memory, and negative target smoothing.
The deployed configuration and its serving costs are
discussed in Section~\ref{sec:production_deployment}.

The teacher provides a quality reference using the retained lifetime
history without compression. Serving it would require processing extensive
event sequences and keeping those sequences accessible in runtime user
profiles. The resulting inference workload and profile storage costs make
this configuration impractical under our production resource and latency
constraints. X-split ARGUS uses precomputed history states and a bounded
recent history during online ranking.

\paragraph{Normalized entropy.}
We use normalized entropy (NE), a standard metric for assessing the quality
of predicted click probabilities~\cite{practicalctr2014}. Let $\mathcal E$
denote the evaluated impressions, $y_i$ the observed binary click feedback,
and $\ell_i$ the binary cross-entropy of the predicted probability for
impression $i$. Sampling correction is applied through impression weights
$w_i$, which are used in both the test log loss and the background CTR $p_0$:
\begin{equation}
    \begin{aligned}
        \mathcal L_{\mathrm{test}}
        &=\frac{\sum_{i\in\mathcal E} w_i\ell_i}
        {\sum_{i\in\mathcal E} w_i},\\
        p_0
        &=\frac{\sum_{i\in\mathcal E} w_i y_i}
        {\sum_{i\in\mathcal E} w_i}.
    \end{aligned}
    \label{eq:weighted_test_statistics}
\end{equation}
Both quantities are computed over the entire test set. NE normalizes the
weighted log loss by the binary entropy of this background CTR:
\begin{equation}
    \operatorname{NE}
    =\frac{\mathcal L_{\mathrm{test}}}
    {-p_0\log p_0-(1-p_0)\log(1-p_0)}.
    \label{eq:normalized_entropy}
\end{equation}
Lower NE indicates better prediction quality. For a model with NE $v$ and
a reference with NE $v_{\mathrm{ref}}$, we report the relative reduction
\begin{equation}
    \Delta\operatorname{NE}(v;v_{\mathrm{ref}})
    =100\frac{v_{\mathrm{ref}}-v}{v_{\mathrm{ref}}}\%.
    \label{eq:relative_ne_reduction}
\end{equation}
A positive reduction denotes an improvement. Each table explicitly identifies
its reference. NE is displayed to four decimal places; relative reductions
and retained teacher gains are computed from unrounded values.
On this test set, a relative NE change of 0.1\% is statistically significant.

\subsection{Baseline}
\label{sec:offline_baseline}

Context-aware ARGUS is a two-tower ranker based on ARGUS~\cite{argus}
whose user tower contains a history compressor and a runtime encoder,
while its item tower represents the candidate ad. It scores ads with a
dot product between the two tower outputs and therefore already models
both historical context and recent behavior. The \emph{Context-aware}
variant separates each event into context and item components. This
separation allows the runtime user encoder to receive all available
contextual information when scoring the corresponding item. At serving
time, context enters the user encoder while the candidate item is
represented by the item tower. The baseline therefore combines historical
behavior with the context available at ranking time.

\subsection{Comparison with the Baseline and Teacher}
\label{sec:offline_main}

Table~\ref{tab:offline_main} shows that X-split ARGUS improves NE over
Context-aware ARGUS, while the uncompressed teacher achieves the lowest NE.
The result supports the combined design of target attention and teacher
supervision for ranking with compressed history.

\begin{table}[tbp]
    \centering
    \caption{Overall comparison of complete model configurations on the
    common RTB test set. Relative reductions use Context-aware ARGUS as
    the reference. $\rho$ is the share of the teacher's NE improvement
    retained, defined in Equation~\ref{eq:retained_teacher_gain}.}
    \label{tab:offline_main}
    \small
    \setlength{\tabcolsep}{4pt}
    \begin{tabularx}{\linewidth}{@{}>{\raggedright\arraybackslash}Xrrr@{}}
        \toprule
        Model & NE $\downarrow$ & \shortstack{$\Delta$NE\\(\%) $\uparrow$} & \shortstack{$\rho$\\(\%) $\uparrow$} \\
        \midrule
        Context-aware ARGUS & 0.7204 & 0.000 & --- \\
        X-split ARGUS & 0.7146 & +0.814 & 71.6 \\
        Teacher & 0.7123 & +1.136 & 100.0 \\
        \bottomrule
    \end{tabularx}
\end{table}

To quantify how much of the teacher's improvement a student retains, we
compare their gains over the same Context-aware ARGUS baseline:
\begin{equation}
    \rho(v)
    =100\frac{v_{\mathrm{base}}-v}
    {v_{\mathrm{base}}-v_{\mathrm T}}\%,
    \label{eq:retained_teacher_gain}
\end{equation}
where $v_{\mathrm{base}}$ and $v_{\mathrm T}$ are the NE of the baseline
and teacher, respectively. X-split ARGUS retains more than 70\% of the
teacher's improvement. This shows that substantial predictive benefit
can be retained in a student that uses compact history representations
and does not execute the teacher using the full history during serving.

\subsection{Ablation Study}
\label{sec:offline_ablations}

To complement the overall comparison, we study how the X-split ARGUS baseline
responds to changes in compressed and recent context, encoder depth, and
negative training targets. Table~\ref{tab:offline_architecture} reports
changes to the compressed history budget $K$, runtime history budget $M$,
and encoder depth $L$ relative to this ablation baseline.
In the history budget experiments, the split between the compressed prefix
and recent history is fixed at its reference value: we vary $K$ or $M$
independently, without moving the split or changing other hyperparameters.
In the depth experiments, both the compressor and the runtime encoder have
$L$ layers, while the runtime attention window at each layer remains fixed.
The smoothing comparison holds $K=2$ in both configurations.

\begin{table}[tbp]
    \centering
    \caption{Configuration sensitivity relative to the X-split ARGUS
    baseline defined in Table~\ref{tab:offline_baseline_configuration}
    in Appendix~\ref{app:design_configuration}.
    The history split is fixed across token budget experiments; the runtime
    attention window at each layer is fixed across depth experiments.
    Negative reductions denote higher NE than the ablation baseline.}
    \label{tab:offline_architecture}
    \small
    \setlength{\tabcolsep}{5pt}
    \begin{tabularx}{\linewidth}{@{}>{\raggedright\arraybackslash}Xrr@{}}
        \toprule
        Configuration & NE $\downarrow$ & $\Delta$NE (\%) $\uparrow$ \\
        \midrule
        X-split ARGUS baseline & 0.7155 & 0.000 \\
        \addlinespace
        \multicolumn{3}{@{}l}{\emph{Compressed history tokens}} \\
        $K=2$ & 0.7168 & $-0.193$ \\
        $K=32$ & 0.7149 & $+0.075$ \\
        \addlinespace
        \multicolumn{3}{@{}l}{\emph{Runtime history states}} \\
        $M=8$ & 0.7179 & $-0.342$ \\
        $M=64$ & 0.7157 & $-0.029$ \\
        \addlinespace
        \multicolumn{3}{@{}l}{\emph{Layers in each history encoder}} \\
        $L=8$ & 0.7148 & $+0.091$ \\
        $L=10$ & 0.7137 & $+0.245$ \\
        \bottomrule
    \end{tabularx}
\end{table}

\paragraph{Compressed history tokens.}
We compare compressed token budgets of $K=2$, $8$, and $32$. Increasing
the number of compressor states available to candidate attention improves
prediction quality across these settings. Reducing the budget below the
ablation baseline causes a larger quality loss than the gain from
increasing it above the baseline.

\paragraph{Runtime history tokens.}
We compare runtime history budgets of $M=8$, $64$, and $256$ events.
For each target and budget, we recompute the runtime encoder on the last
$M$ raw events and pass their output states to the decoder. Earlier events
are excluded from this encoder pass, while the compressed prefix remains
the same across configurations. The smallest budget produces the largest
loss in prediction quality, while the intermediate budget is closer to
the ablation baseline. Retaining more recent events in the runtime branch
therefore improves prediction quality with the same compressed historical
context.

\paragraph{Encoder depth.}
We increase both history encoders from six to eight and ten layers.
Deeper encoders improve prediction quality across the tested
configurations without increasing the number of history tokens exposed to
the decoder. With a fixed window at each layer,
additional layers can also expand the context propagated through the
runtime encoder stack. The experiment therefore measures the combined
effect of depth in both branches, without separating their individual
contributions or the effects of capacity and context propagation.

\paragraph{Smoothing negative targets.}
\label{sec:offline_smoothing}
We test whether regularizing negative training targets with the asymmetric
smoothing defined in Section~\ref{sec:design_smoothing} improves prediction
quality. The paired comparison uses the same $K=2$ configuration with
binary or smoothed targets, keeping all other settings unchanged.

\begin{table}[tbp]
    \centering
    \caption{Negative target smoothing at fixed $K=2$. Relative NE
    reductions use the configuration with binary targets as the reference.}
    \label{tab:offline_smoothing}
    \small
    \setlength{\tabcolsep}{4pt}
    \begin{tabularx}{\linewidth}{@{}>{\raggedright\arraybackslash}Xrr@{}}
        \toprule
        Targets & NE $\downarrow$ & \shortstack{$\Delta$NE (\%)\\vs. binary} \\
        \midrule
        Binary & 0.7168 & 0.000 \\
        Smoothed negatives & 0.7155 & $+0.179$ \\
        \bottomrule
    \end{tabularx}
\end{table}

Table~\ref{tab:offline_smoothing} shows that smoothing produces a
statistically significant improvement in prediction quality over binary
supervision without changing training or online inference cost.

\section{Production Deployment}
\label{sec:production_deployment}

We deploy X-split ARGUS on the largest advertising surface of our real-time
bidding system. This section reports its online impact, the resources
required to serve it, and the processes used to refresh historical
representations and model parameters. The deployed model configuration is
listed in Table~\ref{tab:design_configuration} in
Appendix~\ref{app:design_configuration}.

\subsection{Online Experiment and Deployment Impact}
\label{sec:production_online}

The surface accounts for approximately 50\% of total RTB traffic and
handles roughly a hundred thousand requests per second. We conducted an
A/B test lasting seven days against the prior production ranker, Context-aware ARGUS
(Section~\ref{sec:offline_baseline}). Users were assigned to treatment and
control by hashing their identifiers with an experiment salt, keeping
assignment consistent across requests. Each group received 6\% of the
surface's traffic; the request rate above refers to the entire surface.

Table~\ref{tab:production_online} reports relative changes against the
control. X-split ARGUS increases ACV by \textbf{4.65\%} and the number of
clicks by \textbf{2.96\%}. ACV is our primary internal metric for advertising
A/B acceptance and captures the monetary value of advertiser conversions
delivered at the same advertising budget.
Both improvements are statistically significant
($p<0.001$).

A reverse
A/B test conducted one month after deployment confirmed the gains.

\begin{table}[tbp]
    \centering
    \caption{Online A/B results relative to the Context-aware ARGUS
    production baseline. Each group received 6\% of the evaluated
    surface's traffic during the test, which lasted seven days.}
    \label{tab:production_online}
    \small
    \setlength{\tabcolsep}{5pt}
    \begin{tabularx}{\linewidth}{@{}>{\raggedright\arraybackslash}Xr@{}}
        \toprule
        Metric & Relative change \\
        \midrule
        ACV & \textbf{+4.65\%} \\
        Number of clicks & \textbf{+2.96\%} \\
        \bottomrule
    \end{tabularx}
\end{table}

At this request volume, deployment must satisfy both the latency budget
imposed by user response requirements and the throughput supported by
the available GPUs. The ML service has a latency budget in the hundreds
of milliseconds. X-split ARGUS meets the same budget as Context-aware
ARGUS while requiring 15\% more online GPUs of the same type to serve the
same traffic volume. The computation reuse described in
Section~\ref{sec:design_student} keeps long history processing outside the
request path and shares recent history encoding across candidates. The
online gains are thus achieved within the existing latency constraint,
with a measured increase in online GPU capacity. Precomputation, storage,
and model updates incur additional resources, described below.

\subsection{History Precomputation and Storage}
\label{sec:production_precomputation}

A daily batch pipeline collects new user events, appends them to the
retained user profiles, and runs the history compressor for users with
new events. For each such user, it publishes $K$ contextualized history
states to an embedding store. The online ranker retrieves these states
and combines them with representations of recent events, as described
in Section~\ref{sec:design_student}. Historical representations are
therefore reused across requests, while recent behavior remains available
through the runtime user profile.

We store the compressor states in BF16. For a stored state dimension
$d_{\mathrm c}$, the uncompressed embedding payload per user is
\begin{equation}
    S_{\mathrm{user}}=2K d_{\mathrm c}\ \text{bytes}.
    \label{eq:production_embedding_storage}
\end{equation}
With standard storage compression, the database of precomputed embeddings
occupies hundreds of gigabytes. This footprint refers to the computed
embeddings, separately from the retained event histories used by the
batch pipeline. The GPU compute required for daily precomputation is
small relative to continuous online inference; storing the resulting
representations is an additional infrastructure requirement.

\subsection{Model Freshness and Daily Fine-Tuning}
\label{sec:production_freshness}

\begin{figure}[tbp]
    \centering
    \includegraphics[width=\linewidth]{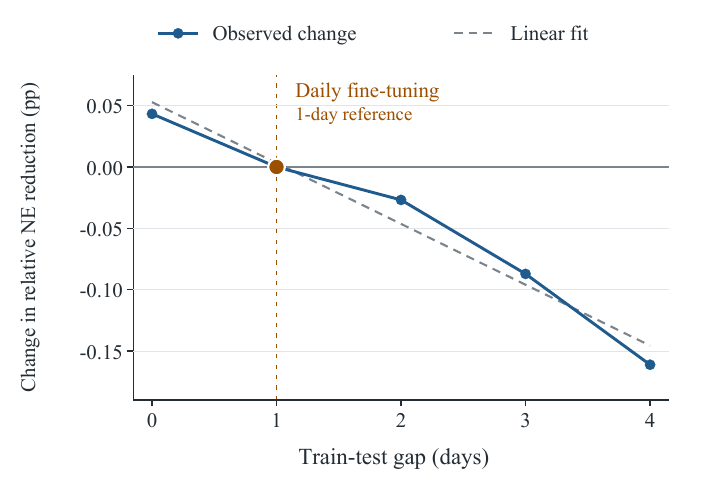}
    \caption{Relative prediction quality deteriorates as the gap between the latest
    training event and the test day increases. Histories and compressor
    embeddings are refreshed while model weights remain fixed.}
    \Description{The model's relative advantage over production declines approximately linearly as the gap between its latest training event and the test day increases from zero to four days. Histories are refreshed while model weights remain fixed.}
    \label{fig:production_freshness}
\end{figure}

To inform the update schedule, we evaluate a fixed X-split ARGUS checkpoint
on test windows spanning one day, with gaps of zero to four days between
the latest training event and the start of each window. User histories
and precomputed compressor embeddings are refreshed using events available
before each prediction. Figure~\ref{fig:production_freshness} shows the
change in relative NE reduction over the production system evaluated on
each test day, with the value at a gap of one day set to zero. This advantage decreases
approximately linearly, by about 0.05 percentage points per day,
despite the refreshed historical context.

Based on this observation, we fine-tune the deployed student daily using
prediction targets from the most recent three days. Data preparation,
training, and deployment yield a gap of approximately one to two days
between the latest training events and serving predictions.
Daily fine-tuning requires additional GPU compute,
but this workload is small relative to continuous online inference.
The larger improvement with no gap suggests that more frequent or
online model updates could yield further gains.

\section{Conclusion}
\label{sec:conclusion}

We presented X-split ARGUS, a framework for CTR prediction that combines
compressed historical context with recent user behavior and candidate
scoring through cross-at\-ten\-tion. Its history compressor and online ranker
are trained end to end using observed click feedback and distillation
from a teacher that processes the uncompressed history. Precomputed
compressor states are reused across requests, while runtime history
representations are shared across candidates, keeping extensive history
processing outside the online ranking path. In the offline comparison,
the compressed student improves normalized entropy over Context-aware
ARGUS and retains more than 70\% of the uncompressed teacher's improvement
over the same baseline.

Deployment on the largest advertising surface of our RTB system confirms
the practical value of this design. An online A/B test against the
production baseline yields relative gains of \textbf{4.65\%} in ACV and \textbf{2.96\%} in
clicks. X-split ARGUS serves the same traffic volume within
the existing latency budget using 15\% more online GPUs of the same type.
Daily history precomputation and model fine-tuning support continued
operation, with the latter motivated by the observed decline in the
relative advantage of a fixed checkpoint. These results demonstrate that
reusable history compression and candidate attention can bring the
benefits of richer user context to production ranking within practical
latency and resource constraints.

\bibliographystyle{unsrtnat}
\bibliography{references}

\clearpage
\appendix

\section{X-split ARGUS Configuration}
\label{app:design_configuration}

Table~\ref{tab:design_configuration} reports the model configuration and loss
weights used in the deployment described in this paper. Sequence lengths
count original events, before the teacher represents target events as
separate item and feedback tokens.
The maximum history length corresponds to the 99.5th percentile of the
distribution of event counts.
The deployed runtime encoder uses a window parameter $w=40$ at each of its
$L_{\mathrm r}=6$ layers and a configured recent history budget
$M=wL_{\mathrm r}=240$.

\begin{table}[htbp]
    \centering
    \caption{Model and training hyperparameters.}
    \label{tab:design_configuration}
    \small
    \begin{tabularx}{\linewidth}{@{}>{\raggedright\arraybackslash}Xr@{}}
        \toprule
        Hyperparameter & Value \\
        \midrule
        Maximum history horizon & 1 year \\
        Maximum history length $N$ & 8192 \\
        Recent history budget $M$ & 240 \\
        Compressed history tokens $K$ & 8 \\
        Compressor layers & 6 \\
        Compressor hidden dimension & 1024 \\
        Runtime encoder layers $L_{\mathrm r}$ & 6 \\
        Runtime attention window $w$ & 40 \\
        Runtime encoder hidden dimension & 1024 \\
        Decoder layers & 2 \\
        Decoder hidden dimension & 256 \\
        Distillation weight $\lambda_{\mathrm{KD}}$ & 0.15 \\
        Smoothed negative target $\varepsilon$ & 0.01 \\
        \bottomrule
    \end{tabularx}
\end{table}

Table~\ref{tab:offline_baseline_configuration} defines the X-split ARGUS
baseline used for the offline ablation study. The X-split ARGUS
configuration in the main offline comparison additionally enables
self-at\-ten\-tion over history and negative target smoothing.

\begin{table}[htbp]
    \centering
    \caption{X-split ARGUS baseline settings for the offline ablation study.}
    \label{tab:offline_baseline_configuration}
    \small
    \begin{tabularx}{\linewidth}{@{}>{\raggedright\arraybackslash}Xr@{}}
        \toprule
        Setting & Value \\
        \midrule
        Compressed history tokens $K$ & 8 \\
        Recent history budget $M$ & 256 \\
        Compressor layers & 6 \\
        Runtime encoder layers & 6 \\
        Decoder self-at\-ten\-tion over history & Disabled \\
        Supervised targets & Binary \\
        Knowledge distillation & Enabled \\
        \bottomrule
    \end{tabularx}
\end{table}

For the X-split ARGUS baseline, the runtime encoder uses a window
parameter at each layer of
$w=\lfloor S/L_{\mathrm r}\rfloor$, where $S$ is the configured visibility
budget (\texttt{seqlen}). In the depth experiments of
Section~\ref{sec:offline_ablations}, this window remains fixed as the number
of layers increases; it is not recomputed for each depth.
The window bounds the context propagated through the encoder stack.
It does not make the last $M$ states of a full training
sequence identical to states recomputed from only the last $M$ raw events
during serving, as discussed in Section~\ref{sec:design_training}.

\end{document}